# A combined experimental and numerical study of stab-penetration forces

Aisling (first name) Ní Annaidh (surname)[a], Marie Cassidy[b], Michael Curtis[b], Michel Destrade[a,c], Michael D. Gilchrist[a,*]

[a]*School of Mechanical & Materials Engineering, University College Dublin, Belfield, Dublin 4, Ireland*
[b]*Office of the State Pathologist, Fire Brigade Training Centre, Malahide Road, Marino, Dublin 3, Ireland*
[c]*School of Mathematics, Statistics and Applied Mathematics, National University of Ireland Galway, Galway, Ireland*

---

[*]Corresponding author: Phone: +353 1 716 1890
*Email addresses:* `aisling.niannaidh@ucd.ie` (Aisling (first name) Ní Annaidh (surname)), `mcassidy@statepathology.gov.ie` (Marie Cassidy), `mcurtis@statepathology.gov.ie` (Michael Curtis), `michel.destrade@nuigalway.ie` (Michel Destrade), `michael.gilchrist@ucd.ie` (Michael D. Gilchrist )

# A combined experimental and numerical study of stab-penetration forces

**Abstract**

The magnitude of force used in a stabbing incident can be difficult to quantify, although the estimate given by forensic pathologists is often seen as 'critical' evidence in medico-legal situations. The main objective of this study is to develop a quantitative measure of the force associated with a knife stabbing biological tissue, using a combined experimental and numerical technique. A series of stab-penetration tests were performed to quantify the force required for a blade to penetrate skin at various speeds and using different 'sharp' instruments. A computational model of blade penetration was developed using ABAQUS/EXPLICIT, a non-linear finite element analysis (FEA) commercial package. This model, which incorporated element deletion along with a suitable failure criterion, is capable of systematically quantifying the effect of the many variables affecting a stab event. This quantitative data could, in time, lead to the development of a predictive model that could help indicate the level of force used in a particular stabbing incident.

## 1. Introduction

When a stabbing is fatal, the amount of force required to inflict the stab wound is often the source of much debate in court. It is, of course, in the interest of the prosecution to claim that the level of force was severe or frenzied, and conversely the defense would rather have the force described as mild or benign. It is inevitable that the forensic pathologist, as an expert witness, will be asked to quantify the force involved in the stabbing attack. However, it is impossible to exactly quantify the force exerted, although a qualitative assessment can be made by the pathologist, based on the following four variables [1]:

- The condition of the knife. Is the blade sharp or blunt?

- The resistance offered by different tissues. Was cartilage or bone damaged?

- The depth and length of wound. i.e. if a 2 inch blade causes a 4 inch deep stab wound, considerable force must have been applied to result in the compression of the tissue.

- The amount of clothing present on the victim and its composition. The clothing provides an additional protective layer to penetration.

Based on this assessment the answer given by the expert witness will be a qualitative description using comparative adjectives such as 'mild', 'moderate', 'considerable' or 'severe'. The problem with such descriptions is that their interpretation is highly subjective. For example, an expert witness might consider a particular force to be moderate and a juror may consider the same force to be severe.

While the problem described here is unique to forensic pathology, it is clear that it could benefit from an inter-disciplinary approach incorporating biomechanics. In recent years, the fields of biomechanics and forensic medicine have merged to form a new discipline, forensic biomechanics. This discipline has met the needs of the legal system in particular, with biomechanists increasingly acting as expert witnesses in a court of law [2]. Previous studies on the topic of stab penetration have focused on experimental testing alone. Typically these studies use customized instrumented knives, where the stab-penetration test is performed by a volunteer holding the blade [3, 4, 5, 6]. An alternative technique, which is used here, is to attach an instrumented blade to a machine which drives the stabbing motion [7, 8], offering greater control over a range of test variables. In this study, the data obtained through experimentation has been used to develop a finite element model of stab penetration. Finite element analysis (FEA), originally developed in the 1940s as a tool for civil and aeronautical engineering, has become an invaluable tool in biomechanics over the last three decades [9]. With the exception of [7], who carried out a preliminary study to investigate the suitability of FEA to simulate stab penetration tests, to the best of the author's knowledge, FEA has not yet been used as a tool in the investigation of stab mechanics. The model developed here, is therefore the first fully developed FEA model which simulates the penetration of a blade into human skin. The model replicates the conditions of the stab-penetration test and uses the Von Mises stress criterion coupled with element deletion to model the failure of the skin. The chief advantage of developing such a model is that once the development process is complete, the model can be used to investigate the influence of the many parameters associated with stabbing incidents.

## 2. Materials and methods

*2.1. Experimental*

Stab-penetration tests were performed on porcine skin and polyurethane and a limited number of tests were performed on human skin. The skin was cut into a cruciform shape using a custom made die. Open-cell polyethylene foam of density $35\text{kg/m}^3$ was placed below the target material to ensure that it did not deform nor vibrate excessively, which would not be representative of a real-life scenario. Open-cell polyethylene foam has previously been found to be a suitable surrogate material [10, 11, 8] and this density is close to that used by [10] and by [11].

Experiments were carried out at a range of test speeds from 100 mm/min (quasi-static) to 9.2 m/s, consistent with the typical maximum velocity of the arm in a stabbing motion [12, 10]. The experimental set-up for the stab-penetration tests requires three main devices: The biaxial tension device, whose primary function is to hold the test material in place, the blade holder, whose function is to secure the blade or other implement in place on the test machine, and the test machine itself. In the case of quasi-static tests, a Tinius Olsen universal testing machine, and in the case of dynamic tests a Rosand droptower was used. The biaxial tension device and blade holder are illustrated in Fig. 1 [8]. The device is designed in such a way that the test material can be held in biaxial tension by adjusting the lead screws of the clamping mechanism. The blade holder design allows for easy interchange of blades for testing.

The knives most commonly used in stabbing incidents are those household knives which are most readily available [13]. Three knives commonly available in the household have been tested; a cook's knife, a carving knife and a utility knife (shown in Fig. 2). The blade tip geometry, or the sharpness of the blade plays a major role in the amount of force used in stabbing incidents. For this reason it is important to capture the geometry of the blade accurately, in particular for the modeling of such geometries in a finite element analysis study. The blades were characterized by the following indicative dimensions; blade tip angle, blade tip radius, cutting edge angle and cross-sectional thickness (shown in Fig. 3). The cutting edge angle was measured using optical microscopes, while the blade tip radius and the blade tip angle were measured using Scanning Electron Microscope (SEM). The stab-penetration forces of three common non-blade implements were also investigated i.e. a closed pair of scissors, a Phillips screwdriver and a flat head screwdriver.

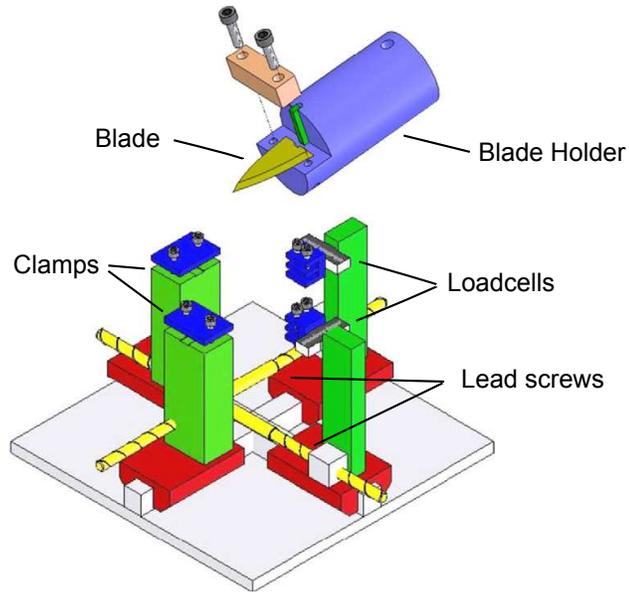

Figure 1: Illustration of biaxial device.

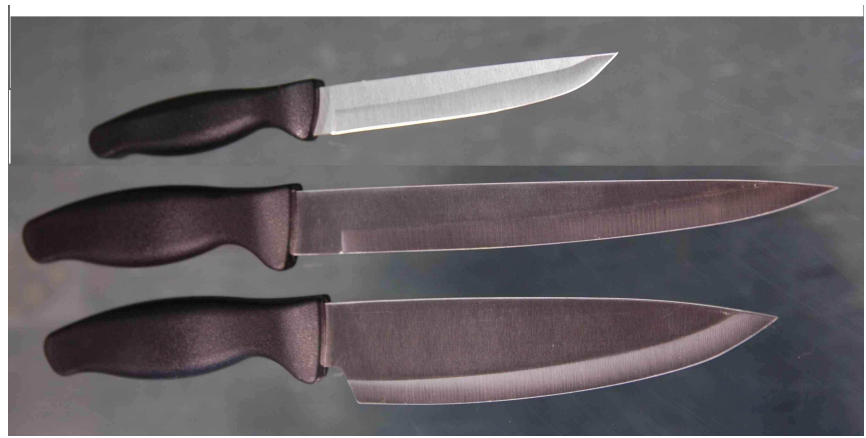

Figure 2: Selection of knives used in experiments. From top: utility knife, carving knife, cook's knife.

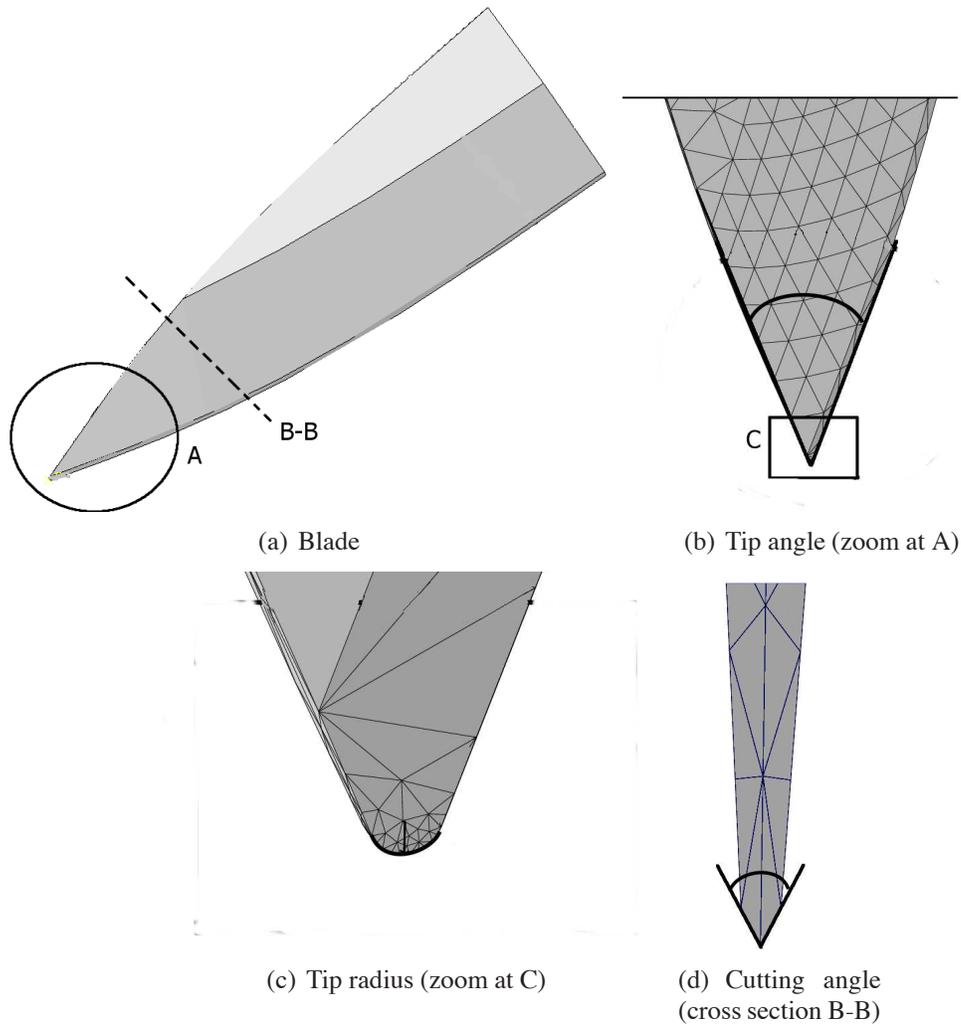

(a) Blade

(b) Tip angle (zoom at A)

(c) Tip radius (zoom at C)

(d) Cutting angle (cross section B-B)

Figure 3: Indicative blade dimensions.

*2.2. Finite element analysis*

The accuracy of an FEA model is heavily dependent upon the accuracy and suitability of its material definitions. The Gasser-Ogden-Holzapfel (GOH) model is a popular structurally based strain energy potential that is commonly used to model the behavior of arteries [14, 15]. The GOH model has been chosen here to model the behavior of skin. The material definitions used here have been evaluated directly using a combination of experimental testing and histological investigation of in vitro human skin in [16] and are provided in Table 1.

The failure mechanism employed is one of element deletion. In this method, once the stress in an individual element exceeds a specified threshold, the element is deemed to have failed and is deleted from the model. Here, the failure criterion is met when an element exceeds a Von Mises stress of 21 MPa, which corresponds to the ultimate tensile strength of human skin given in [17].

Table 1: Material parameters of GOH model for skin (evaluated in [16]).

| $\mu$ MPa | $k_1$ MPa | $k_2$ | $\gamma$ ° | $\kappa$ |
|---|---|---|---|---|
| 0.2014 | 24.53 | 0.1327 | 41 | 0.1535 |

A further important aspect of the FEA model is the contact definition i.e. how the materials in contact with eachother will behave. The hard kinematic contact algorithm was chosen with finite sliding contact because it typically performs better when a hard surface, i.e. blade, contacts a much softer one, i.e. skin. The friction coefficient, $\mu$, was chosen as 0.42 based on the findings of [18].

The geometry used to model the stab-penetration tests is shown in Fig.4. Due to mirror symmetry, only one half of the cruciform, blade and substrate was modeled (the symmetry plane is in light gray in Fig. 4). A mesh convergence study was carried out to ensure that the chosen mesh was independent of further increases in mesh density. This check is particularly important for simulations which include element deletion, as element deletion is known to be susceptible to mesh dependency. It was found that at 8000 elements the solution was independent of further increases in the mesh density and this density using C3D8R elements was then chosen for use in all subsequent simulations.[1]

## 3. Results

*3.1. Target materials*

The number of tests performed on human skin was limited. Therefore these tests can only be used to compare against the surrogate materials used, and to validate the developed FE model. Fig. 5 illustrates a typical force-displacement curve at quasi-static speeds for each of the three target materials: human skin, porcine skin and polyurethane. Examining only the shape of the curves, it can be seen that both the porcine skin and the human skin exhibit a non-linear curve within the range of the test, whereas the polyurethane does not. However, a clear advantage of using polyurethane over porcine skin as a surrogate material is that it eliminates the issue of biological variation.

---

[1] Further details of the FE model are available in [19]

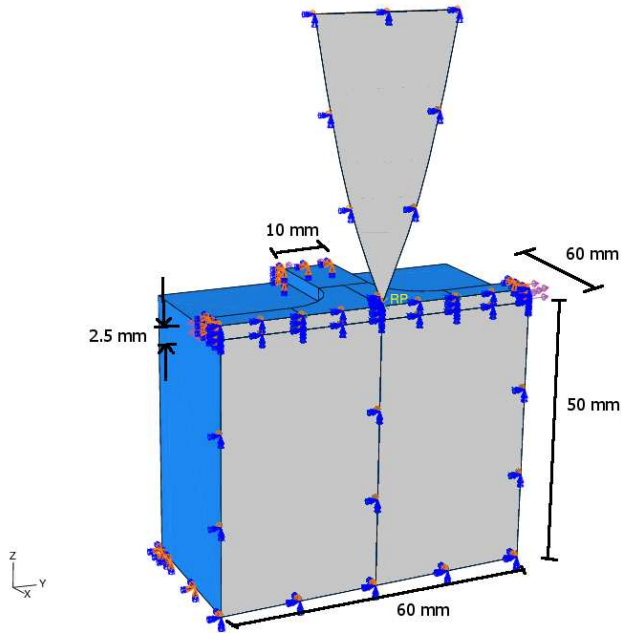

Figure 4: Geometry and boundary conditions of model.

*3.2. Blade geometry*

The three blades used for stab-penetration experiments were modeled using the characteristic blade dimensions given in Fig. 3. Each of these dimensions were quantified experimentally using SEM and microscopy, the results of which are given in Table 2.

Table 2: Summary of blade dimensions.

| Dimension | Cook's | Carving | Utility |
| --- | --- | --- | --- |
| Tip Angle ° | 53 | 35 | 45 |
| Tip Radius $\mu$m | 210 | 125 | 120 |
| Cutting Angle ° | 47 | 50 | 53 |
| Thickness mm | 1.5 | 1.5 | 1 |

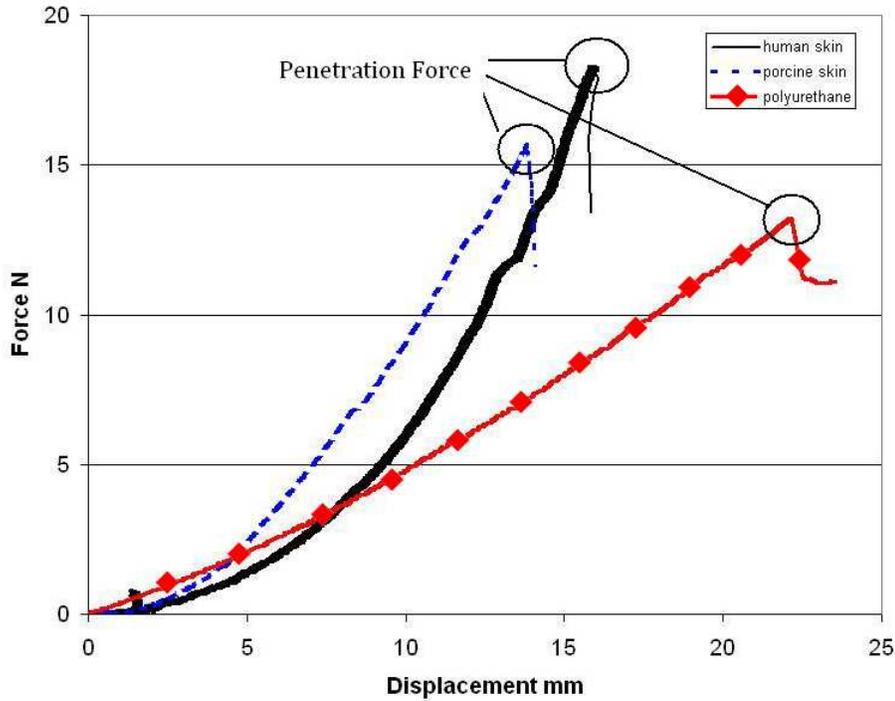

Figure 5: Typical force-displacement curve of a stab-penetration test for each of the target materials at 100mm/min using a carving knife.

*3.3. Presence of clothing*

Stab-penetration tests were carried out on polyurethane at 100 mm/min with various layers of clothing present. Of the four types of clothing tested (cotton, tracksuit, fleece, denim), we see an increase in penetration force varying from 10% for cotton to 50% for denim for a single layer of clothing. Two further tests were carried out with two layers of clothing, namely denim and cotton, and fleece and cotton. It was found that adding additional layers of clothing has an accumulative effect on the penetration force. Fig. 6 provides a graphical summary of these test results.

*3.4. Test speeds*

While at quasi-static speeds the influence of velocity was found to be insignificant in [8], here, where the test speeds were comparable to the maximum and mean velocities attainable during a stabbing incident [12, 10], a clear effect can be seen when increasing the velocity. Fig. 7 compares the mean ($\pm$ standard deviation) maximum penetration force at quasi-static velocities, 1m/s, 4.6m/s and 9.2 m/s, for both polyurethane and porcine skin samples. An analysis of variance (ANOVA) revealed that test speed had a statistically significant effect on the penetration force (P=0.002).

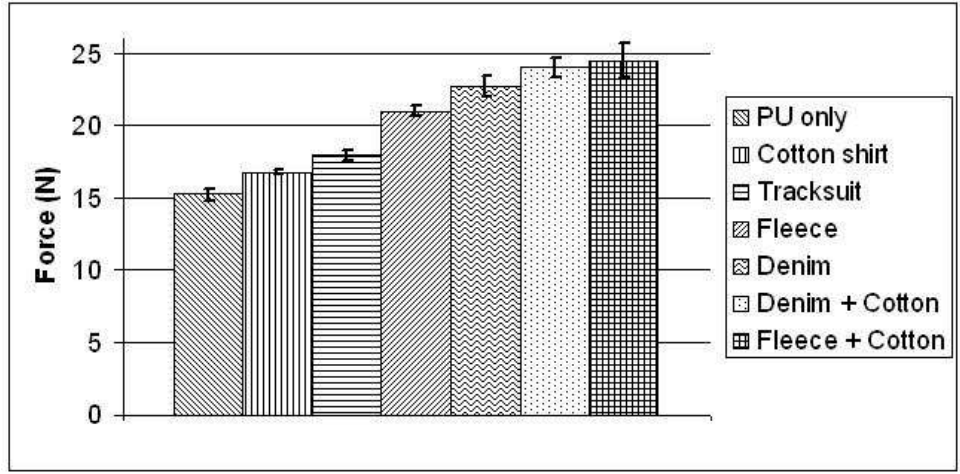

Figure 6: The influence of the presence of various layers of clothing on the maximum penetration force.

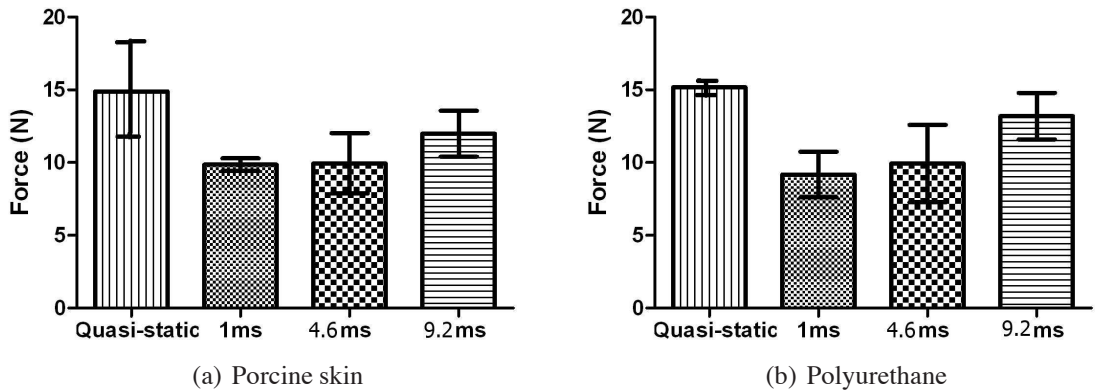

(a) Porcine skin  (b) Polyurethane

Figure 7: Influence of test speed on penetration force of polyurethane and porcine skin using a Cook's knife.

The decrease in penetration force between the quasi-static case and 1 m/s is due to the effects of viscoelasticity whereby the skin behaves stiffer at higher velocities. Since the skin is stiffer at the point of contact with the blade, the energy release rate increases, allowing the crack to propagate through the material more easily, resulting in a reduced rupture force [20]. Interestingly, while the mean penetration force at dynamic speeds was lower than that at quasi-static speeds, when we examine the dynamic speed tests alone, we see that with increased speeds, the force increases. Video analysis suggests that the increase at 9.2 m/s may be due to stress wave propagation through the material upon impact.

*3.5. Implements*

Fig. 8 compares the mean penetration force for each of the implements used, a closed pair of scissors, a flat-head screwdriver, a Phillips screwdriver, a cook's knife, a carving knife and a utility knife. Previous results indicated that the effect of varying the stabbing implement would have a significant effect on the force-displacement curve [3, 4, 8]. Our results support these findings, and in particular those of [8] who found that of the three blades used (cook's, carving and utility knife), the cook's knife had the highest penetration force and the utility knife had the lowest. Comparing the penetration force of the blades, to the non-blade implements, we see that these much blunter objects can exceed 300% the penetration force of a sharp blade. A further interesting result here is that the non-blade implements (scissors, flat head screwdriver, Phillips screwdriver) failed to puncture the skin or polyurethane at 1m/s, and simply rebounded off with no visible damage to the material.

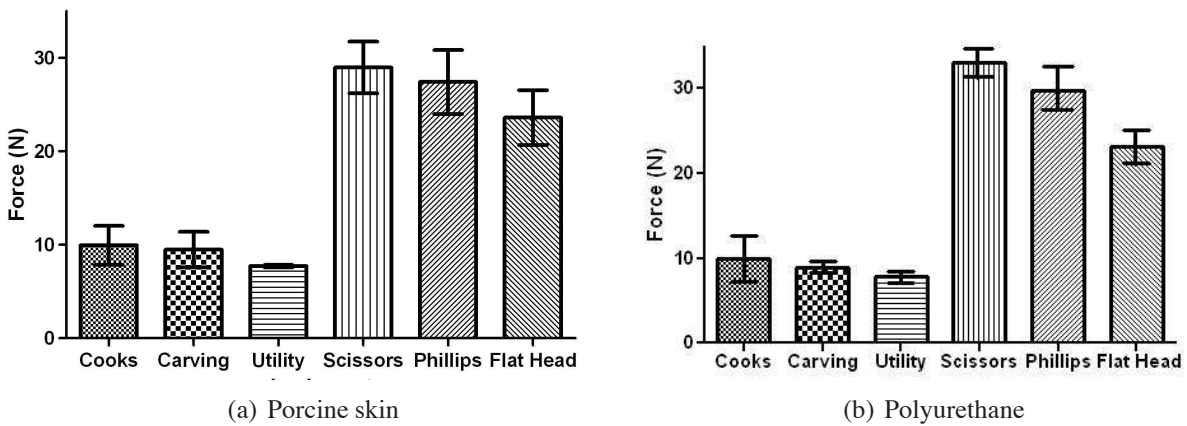

(a) Porcine skin   (b) Polyurethane

Figure 8: Maximum force of various implements at 4.6 m/s.

*3.6. FE simulations*

The element deletion process is shown in detail in Fig.9. In Fig.9 (a), the blade has deformed the skin but no initial penetration has occurred. This corresponds to the early portion of the force-displacement curve shown in Fig.10. Once the threshold value of the failure criterion is reached, the top layer of elements is deleted and the blade then progresses to the next layer as shown in Fig.9(b). The deletion of the top layer corresponds to a small drop in force-displacement curve in Fig.10 (point B and C) until contact resumes again with the blade. This does not affect the maximum penetration force but leads to small intermittent decreases along the force displacement curve until rupture occurs. The blade progresses through the skin in this manner until full perforation of the skin occurs in Fig.9(d), which corresponds to the maximum force shown at point D in Fig.10.

Examining in more detail the results of the FE model simulation, Fig.10 compares the force displacement graph predicted by the FE model against the corresponding experimental graph at quasi-static speeds for human skin. It can be seen that the FEA results offers an excellent match to the experimental data provided here. The maximum penetration force varies by only 1N (6%) and while there is a difference of 2 mm (10%) between the maximum displacement values, this is comparable to the level of variation in the experimental results. It should be noted here that while the FE prediction is excellent at quasi-static test speeds, in its current form, it includes a hyperelastic material model of skin and therefore cannot capture the viscoelastic effects illustrated experimentally in Fig.7.

As further validation, Fig.11 compares the maximum penetration force for each of the three blades and compares this to the experimental values. The magnitude obtained through simulation differs from the experimental mean, but it should be noted that the experimental results here are from tests on polyurethane and the numerical results are from human skin. The important point about this comparison, which is not like for like, is that the simulation results respect the experimental order of magnitudes, with the cook's knife requiring the maximum force, followed by the carving knife and finally, the utility knife.

## 4. Discussion

The primary advantage of the present study over older studies which use an instrumented blade, is that the present experiments were carried out in a controlled and repeatable manner. Of course it is difficult to recreate the exact conditions of a knife attack in a laboratory setting, and the stab-

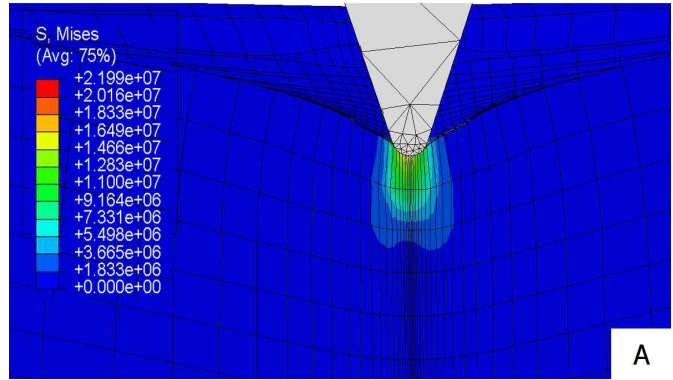
(a)

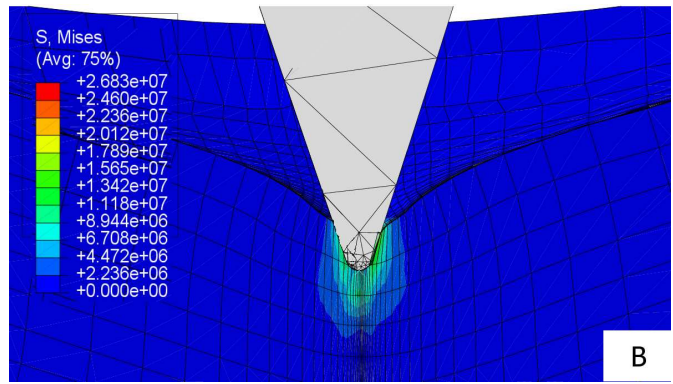
(b)

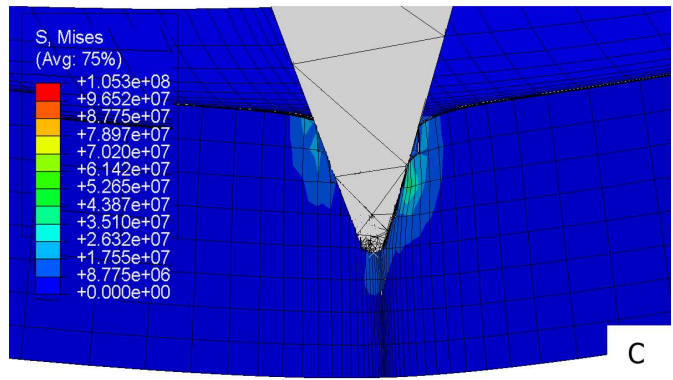
(c)

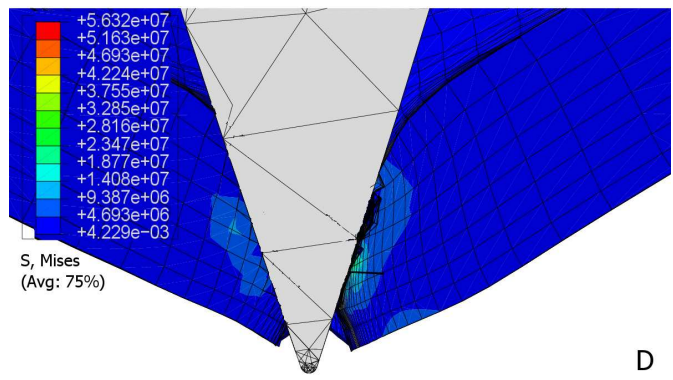
(d)

Figure 9: Numerical progression of a carving knife through human skin (units in Pa).

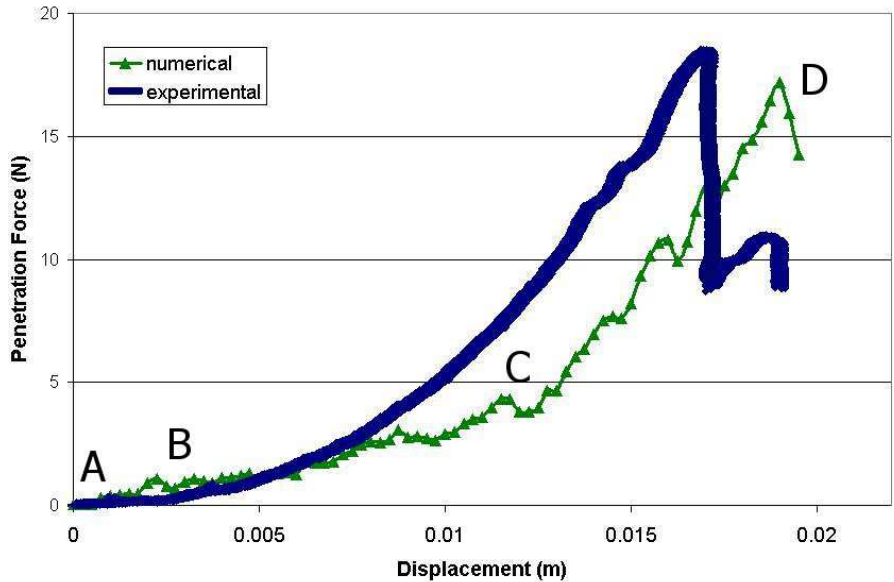

Figure 10: Experimental and numerical force-displacement graph during stab-penetration test of human skin at 100mm/min with a carving knife.

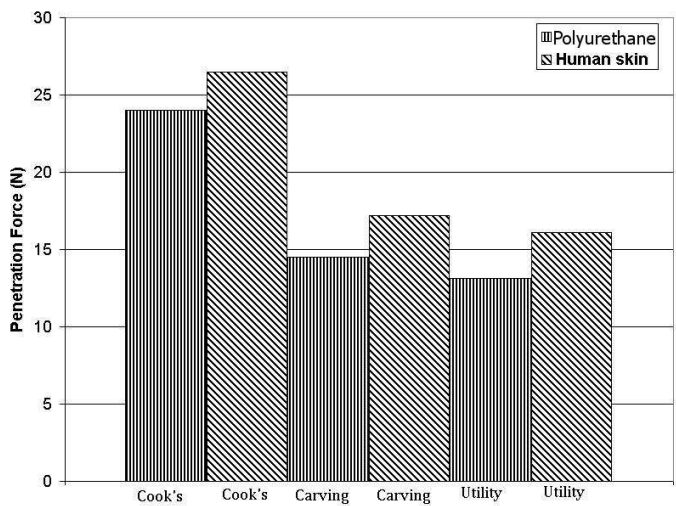

Figure 11: Comparison of experimental (polyurethane) and numerical (human skin) maximum force for different blades at 100mm/min.

penetration test is a simplified model of such an attack. However, methods such as the drop-testing method employed here create good conditions for carrying out comparative studies to examine different blades, different approach speeds etc. [11]. Comparing our own results to those from the literature in Table 3 we see that the present experimental results are similar to those of [3] and [4], who used instrumented blades, and to those of [7] and [8] who used testing machines at quasi-static speeds. Our results appear to differ significantly only from the work of [6], whose forces appear to be substantially greater than all other authors.

Table 3: Summary of stab-penetration (with knives) results from the literature.

| Author | Velocity | Target material | Force |
|---|---|---|---|
| [3] | Quasi-static, Dynamic | Cadaveric tissue | <5N for a sharp blade, 30-50N for a blunt blade |
| [4] | Quasi-static, Dynamic | Cadaveric tissue | <10N for naked skin, 70-100N for clothed |
| [6] | Quasi-static, Dynamic | Cadaveric tissue | 35-55N |
| [7] | Quasi-static | Porcine skin | 10-15N |
| [8] | Quasi-static | Polyurethane | 13-20N |
| Present results | Quasi-static, Dynamic | Cadaveric tissue, Porcine skin, polyurethane | 15-17N at quasi-static, 10-12N at dynamic |

The experimental aspect of this study was performed using human skin, porcine skin and polyurethane. Because the availability of human skin was limited, the majority of tests were performed using surrogate materials. There was, however, sufficient data available to conclude that both porcine skin and polyurethane are suitable surrogate materials for stab-penetration tests. Porcine skin mimics the behavior of human skin closely, having the same non-linear curve, anisotropy, and similar levels of deformation. The emphasis in this study has been placed on the puncture of skin alone, justified by the widely accepted fact that skin offers the largest protection to puncture until bone or cartilage is reached [3, 4, 6]. It is clear however, that the underlying tissue also plays an important role and cannot be ignored completely. For this reason, foam was placed below the target material to model the behavior of underlying soft tissue.

Previous authors have stated that the penetration force decreases as the speed is increased [3, 4]. This effect is partly due to the viscoelastic nature of skin, which appears stiffer at higher velocities, but it is also influenced by fracture mechanics. Increasing the velocity of the blade increases the energy release rate of the skin, which results in a reduced rupture force [20]. This effect is related to the manner of crack propagation and is unique to highly deformable materials [20]. Increasing the velocity of the blade also leads to a reduced contact area between the blade and the skin [21].

Here, it has been shown that the penetration force for dynamic tests overall, are lower than the quasi-static tests, in agreement with [3, 4]. Although it appears that at the 9.2 m/s stress wave propagation in the material leads to a higher penetration force.

It is difficult to provide exact data on the details of clothing worn during stabbing incidents because in many stabbing case files there is no reference to the clothing worn. Nonetheless, a review of 108 stabbings in Ireland between 2000 and 2005 by [22] revealed that in 60 cases clothing was not available for examination, in 13 cases the clothing was examined and in the remaining 35 cases the clothing was examined and reported on in detail. Of these 35 cases it was found that 16 (46%) were wearing only a single layer of clothing, 15 (43%) were wearing two layers and 4 (11%) were wearing 3 layers. It has been shown experimentally that a single layer of clothing can increase the penetration force by up to 50% and future work should aim to include the effects of clothing layers in the FE model.

Previous authors have examined the effect of using various different blades and it is now well established that the blade tip-geometry has a significant effect on the penetration force [3, 4, 8, 23, 24]. [3], in particular, has even examined the effect of using blunted blades. Our results found that such non-blade implements can exceed the penetration force of blades by 300%. Interestingly for forensic investigators, all of the non-blade implements failed to puncture the skin at 1m/s. This places serious doubt over the validity of 'walk-on' style defenses where such non-blade implements have been used. Recently, [25] have quantified the penetration force for other implements such as screwdrivers using silicone gel and related the cross-sectional area of the screwdriver head to the penetration force. Here, we have identified three characteristic dimensions of a blade which affect the penetration force. A further publication [26] will utilize this fact together with the FE model developed here to devise a statistical model of stab-penetration force which could be used to predict the penetration force of a given blade.

## 5. Conclusion

The stab-penetration force of a variety of sharp implements has been experimentally determined using both human skin and suitable surrogates. Comparing the penetration force of the blades to the non-blade implements, it can be seen that these much blunter objects can exceed 300% the penetration force of a sharp blade. Similarly, the influence of the speed of attack has been investigated by examining a range of test speeds varying from quasi-static (50 mm/min) to dynamic (up to 9.2 m/s). An analysis of variance (ANOVA) revealed that test speed had a statistically significant effect on the penetration force (P=0.002). Based on experimental material testing a finite element model has been developed which accurately reflects a stab-penetration event. The model prediction was found to be an excellent match to experimental stab-penetration tests. This FE model has been used in a further publication to develop a statistical model which could predict stab-penetration forces of a given blade.

## 6. Acknowledgments

The authors would like to acknowledge the valuable contribution of Dr. Melanie Otténio and Dr. Karine Bruyère of IFSTTAR (Institut Français des Sciences et Technologies des Transports,de l'Aménagement et des Réseaux) to both tensile testing and a limited number of stab-penetration tests on human skin. This collaborative work has been an essential ingredient in the development of the stab metric.

**References**

[1] D. Dimaio, V. Dimaio, Forensic Pathology, CRC Press, 2001.

[2] M. D. Freeman, S. S. Kohles, Applications and limitations of forensic biomechanics: A bayesian perspective, Journal of Forensic and Legal Medicine 17 (2010) 67–77.

[3] B. Knight, The dynamics of stab wounds, Forensic Science 6 (1975) 249–255.

[4] M. A. Green, Stab wound dynamics–A recording technique for use in medico-legal investigations, Journal of the Forensic Science Society 18 (1978) 161–163.

[5] S. Jones, L. Nokes, S. Leadbeatter, The mechanics of stab wounding, Forensic Science International 67 (1994) 59–63.

[6] P. T. O'Callaghan, M. D. Jones, D. S. James, S. Leadbeatter, C. A. Holt, L. D. M. Nokes, Dynamics of stab wounds: Force required for penetration of various cadaveric human tissues, Forensic Science International 104 (1999) 173–178.

[7] J. Ankersen, A. Birkbeck, R. Thomson, P. Vanezis, Puncture resistance and tensile strength of skin simulants, Proceedings of the Institution of Mechanical Engineers, Part H: Journal of Engineering in Medicine 213 (1999) 493–501.

[8] M. D. Gilchrist, S. Keenan, M. Curtis, M. Cassidy, G. Byrne, M. Destrade, Measuring knife stab penetration into skin simulant using a novel biaxial tension device, Forensic Science International 177 (2008) 52–65.

[9] O. C. Zienkiewicz, The birth of the finite element method and of computational mechanics, International Journal for Numerical Methods in Engineering 60 (2004) 3–10.

[10] E. K. J. Chadwick, A. C. Nicol, J. V. Lane, T. G. F. Gray, Biomechanics of knife stab attacks, Forensic Science International 105 (1999) 35–44.

[11] S. Hainsworth, R. Delaney, G. Rutty, How sharp is sharp? Towards quantification of the sharpness and penetration ability of kitchen knives used in stabbings, International Journal of Legal Medicine 122 (2008) 281–291.

[12] S. A. Miller, M. D. Jones, Kinematics of four methods of stabbing: A preliminary study, Forensic Science International 82 (1996) 183–190.

[13] K. Ormstad, T. Karlsson, L. Enkler, B. Law, J. Rajs, Patterns in sharp force fatalities - a comprehensive forensic medical study., Journal of Forensic Science 31 (1986) 529–542.

[14] T. Gasser, R. W. Ogden, G. Holzapfel, Hyperelastic modelling of arterial layers with distributed collagen fibre orientations, Journal of the Royal Society Interface 3 (2006) 15–35.

[15] G. Holzapfel, R. W. Ogden, Constitutive modelling of arteries, Proceedings of The Royal Society of London A 466 (2010) 1551–1597.

[16] A. Ní Annaidh, K. Bruyère, M. Destrade, M. Gilchrist, C. Maurini, M. Otténio, G. Saccomandi, Automated estimation of collagen fibre dispersion in the dermis and its contribution to the anisotropic behaviour of skin, Annals of Biomedical Engineering 40 (2012) 1666–1678.

[17] A. Ní Annaidh, K. Bruyère, M. Destrade, M. Gilchrist, M. Otténio, Characterising the anisotropic mechanical properties of excised human skin, Journal of the Mechanical Behavior of Biomedical Materials 5 (2012) 139–148.

[18] A. Elkhyat, C. Courderot-MAsuyer, P. Humbert, Influence of the hydrophobic and hydrophilic characteristics of sliding and slider surfaces on friction coefficient: *In vivo* human skin friction comparison, Skin Research and Technology 10 (2004) 215–221.

[19] A. Ní Annaidh, The mechanics of stabbing: A combined experimental and numerical study of sharp force injury, Ph.D. thesis, University College Dublin, 2012.

[20] M. Mahvash, P. E. Dupont, Mechanics of dynamic needle insertion into a biological material, IEEE Transactions on Biomedical Engineering 57 (2010) 934–943.

[21] K. L. Johnson, Contact Mechanics, Cambridge University Press, 1987.

[22] L. Murphy, An investigation into the magnitude of force required for penetration of single and multiple layers of clothing in stab wounds, Master's thesis, University of Strathclyde, 2008.

[23] C. T. McCarthy, M. Hussey, M. D. Gilchrist, On the sharpness of straight edge blades in cutting soft solids: Part I - Indentation experiments, Engineering Fracture Mechanics 74 (2007) 2205–2224.

[24] C. T. McCarthy, A. Ní Annaidh, M. D. Gilchrist, On the sharpness of straight edge blades in cutting soft solids: Part II - Analysis of blade geometry, Engineering Fracture Mechanics 77 (2010) 437–451.

[25] K. Parmar, S. Hainsworth, G. Rutty, Quantification of forces required for stabbing with screwdrivers and other blunter instruments, International Journal of Legal Medicine 126 (2012) 43–53.

[26] A. Ní Annaidh, M. Cassidy, M. Curtis, M. Destrade, M. Gilchrist, Towards a predictive assessment of stab-penetration forces, Submitted (2012).